# An Ultra-high-Speed Waveform Digitizer Prototype Based on Gigabit Ethernet for Plasma Diagnostics

Weigang Yin, Lian Chen, Feng Li, and Ge Jin

*Abstract*—An ultra-high-speed waveform digitizer prototype based on gigabit Ethernet has been developed. The prototype is designed to read out signals of detectors to realize the accurate measurement of various physical quantities for plasma diagnostics. The prototype includes an ultra-high-speed analog-to-digital converter (ADC) used to realize high speed digitization, a Xilinx Kintex-7 field-programmable gate array (FPGA) used for system configuration and digital signal processing, a DDR3 memory bar for data storage, and a gigabit Ethernet transceiver for interfacing with a computer. The sampling rate of the prototype is up to 5Gsps with 10-b resolution. The features of the prototype are described in detail.

## I. Introduction

Signals output from detectors around the plasma contains abundant information about the state of plasma [1]. With the development of the experimental technology, the laser power is gradually improved as well as the number and performance of detectors are increasing [2] [3]. This calls for a growing demand for signal readout systems. Aimed at this demand, we design an ultra-high speed waveform digitizer prototype which can meet the requirements of a variety of measurements such as time, energy and waveform discernment and so on for plasma diagnostics. The prototype includes an ADC with high speed (5 Gsps) and high resolution (10 b), a Xilinx Kintex-7 FPGA used for control and data processing, a DDR3 memory bar for caching the data flow which is generated by ADC at 50Gb/s, and a gigabit Ethernet transceiver working in the user datagram protocol (UDP) for communication with PC. The graphical user interface (GUI) is written based on Labview application platform, achieving the prototype control and data communication. In this paper, the features, structure, and data processing in prototype are described.

## II. Implementation

As shown in Fig. 1, the prototype is composed of analog signal conditioning circuit, ADC, clock circuit, FPGA, DDR3 memory, and gigabit Ethernet transceiver. The core of the prototype is based on an ADC which is equipped with four channels whose sampling rate is 1.25 Gsps each [4]. Four independent analog signal conditioning circuits enable the ADC to work in four-channel mode (1.25Gsps *4), two-channel (2.5Gsps *2) or one-channel mode (5Gsps *1) in order to meet the needs of different measurements. An external trigger input channel enables the prototype to synchronize with the experimental system. The clock circuit provides sampling clock with ultra-low jitter. A DDR3 memory bar with 4GB capacity is employed to cache date from ADC, which makes the storage depth greatly increased, up to 600ms. Data communication is accomplished using gigabit Ethernet. Control of the whole board is done by a Xilinx Kintex-7 FPGA, including prototype configuration, data processing and transmitting.

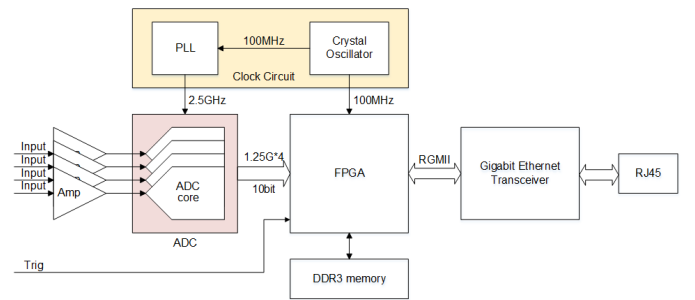

Fig. 1. Schematic of the prototype.

### A. Analog Conditioning Circuit

Protection circuit, single-ended to differential conversion circuit, and anti-aliasing filter form the analog conditioning circuit, which is exhibited in Fig. 2. A TVS diode is placed at the front end of the circuit to protect the circuit from being damaged by electro-static discharge (ESD). The π-type resistor network, which is comprised of R1, R2 and R3, is designed for termination and amplitude adjustment for signal from detector. Diode D2 and D3 is adopted to avoid damage to the circuit caused by large signals. Given the background that most ultra-high-speed ADCs use differential inputs to suppress common-mode noise and interference, a fully differential amplifier is equipped to convert the single-ended detector signal to differential level to match the requirement of ADC. Before the analog signal enters the ADC, an anti-aliasing filter is used to filter out the high frequency noise to improve the signal-to-noise ratio (SNR).

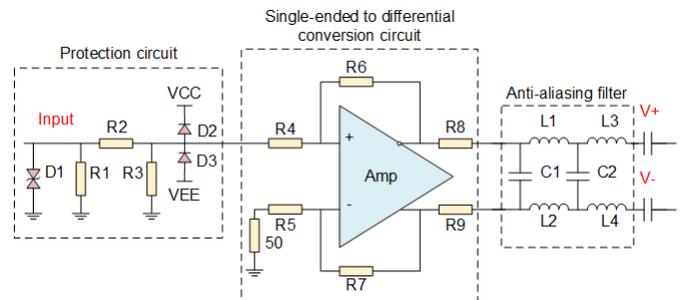

Fig. 2. Analog conditioning circuit.

Weigang Yin is with the State Key Laboratory of Particle Detection and Electronics, University of Science and Technology of China, Hefei 230026, China (e-mail: yd1105@mail.ustc.edu.cn).

Lian Chen is with the State Key Laboratory of Particle Detection and Electronics, University of Science and Technology of China, Hefei 230026, China (email: chenlian@ustc.edu.cn).

Feng Li is with the State Key Laboratory of Particle Detection and Electronics, University of Science and Technology of China, Hefei 230026, China (e-mail: phonelee@ustc.edu.cn).

Ge Jin is with the State Key Laboratory of Particle Detection and Electronics, University of Science and Technology of China, Hefei 230026, China (e-mail: goldjin@ustc.edu.cn).

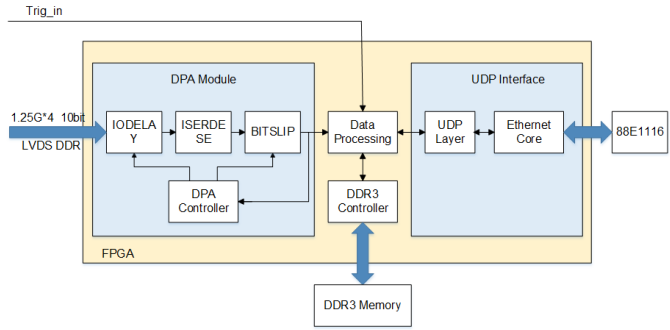

Fig. 3. The signal flow in FPGA of the prototype.

### B. Clock Generation Circuit

In the field of ultra-high-speed analog digital conversion, the jitter of sampling clock have a significant impact on effective number of bits (ENOB) [5], As the frequency of input signal increases, the jitter of the sampling clock will significantly limit the ENOB. The total jitter of sampling is contributed by ADC aperture time itself and the jitter of sampling clock.

In order to achieve high accuracy, an ultra-low jitter phase-locked loop (PLL) is designed to synthesize the sampling clock from the 100MHz clock, which is generated from a precision crystal oscillator. An ordinary crystal oscillator produces another 100MHz clock used for system configuration. Simulation result shows that the jitter of sampling clock is less than 60fs.

### C. Signal Flow in FPGA

The block diagram of the signal flow in FPGA of the prototype is shown in Fig. 3. LVDS interface transmits the digital signal from ADC to FPGA. Then, a dynamic phase alignment (DPA) module is embedded to stabilize the transmission link [6]. When the trigger signal arrives, the signal processing module will transfer the data to DDR3 memory bar through DDR3 controller, which is generated via the core generator of the Xilinx ISE software. After the signal is cached, data will be upload to PC via the gigabit Ethernet.

*1) DPA module*

Since the speed of the interface between the ADC and FPGA reaches 1.25G, the time window of data reception is only about 400 ps. Considering the differences between the line delay and the gate delay of the signal in the chip, coupled with temperature and voltage fluctuations, it is difficult to receive the data accurately and steadily. Therefore, we developed a DPA module based on IODELAY technology for aligning the data signals with the clock signal automatically. The DPA module consists of 2 parts, bit-alignment and word-alignment. The goal of the bit-alignment procedure is to position the captured clock edge in the center of the data eye to provide maximum margin, while the word-alignment procedure aligns the output pattern from the ISERDES to a specific training pattern. Fig. 4 illustrates the state transition diagram of the DPA algorithm.

When the prototype is powered on, it will enter the self-test state. At this stage, the ADC is configured to send a fixed training pattern. Then increase the value of IODELAY in the FPGA to find the left and right edges of the signal. Next, set the IODELAY to the center of the left and right boundaries to get maximum margin. Finally, use BITSLIP to remove word skew and align all channels to a specific word pattern. After DPA is completed, the ADC is switched to normal sampling mode.

*2) UDP interface*

A physical layer chip 88E1116 is used to achieve gigabit Ethernet. The data transfer protocol is based on UDP. Compared with TCP, UDP has higher transmission efficiency and lower resource consumption.

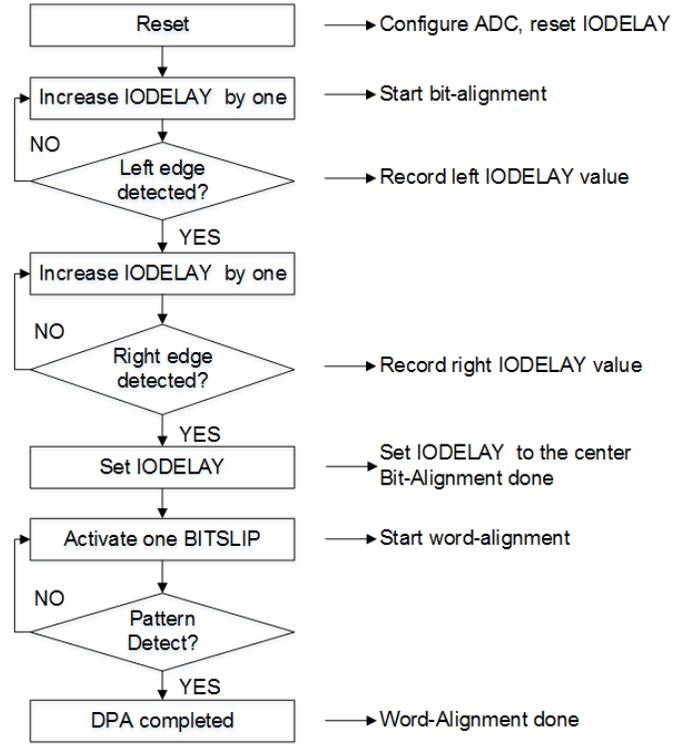

Fig. 4. The state transition diagram of the DPA algorithm.

The data format of an Ethernet packet is shown in Fig. 7(a), it is composed of a 6 bytes destination MAC address, a 6 bytes source MAC address, a 2 bytes packet type, 46-1500 bytes data and a 4 bytes CRC check. When the type is 0x0800, the data which contains 46-1500 bytes is an IP Datagram. The IP Datagram consists of a 20 byte header and a UDP packet. UDP packets include source port, destination port, packet length, and checksum and user data. In this prototype, the length of user data is 1024 bytes per packet.

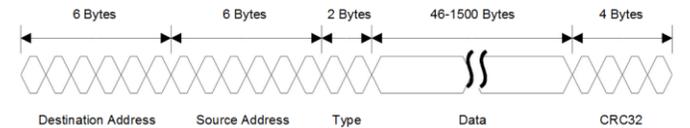

Fig.5. The data format of an Ethernet packet.

## III. TEST

### A. Data Transmission Performances Test

A network test module is built in the FPGA for performance evaluation of the gigabit Ethernet. It can accept the data request command from the host computer, and reply a specific number packets of data to each request. The number of packets ranging from 1 to 256 is set by the data request command. Each packet contains 1024 bytes of user data. A counter for outputting the binary code is used as a data source. These data are checked by the host computer in real time and assess data rate, packet loss rate and code error rate.

The results of the data transmission rate test is shown in Table. I. Each test lasts for one hour. As the N (number of packets sent for each data request command) increases, the transmission speed rises accordingly. When transmitting 256 packets of data each time, the average Ethernet transfer rate reaches up to 813Mb/s, which is less than 1Gb/s. The reasons are mainly attributed to the following. On the one hand, the hardware does not upload Ethernet packet when the host computer sends the data request command. On the other hand, the UDP also occupies part of the bandwidth, which is mainly the tens of bytes of each packet header and the last few bytes of each packet used for CRC check. These reasons decrease the usable bandwidth on a gigabit Ethernet network. Within the 8-hour test, no packet loss and error code are detected. This indicates that the data interface has superior stability.

TABLE I
UNITS FOR MAGNETIC PROPERTIES

| N | Data rate ( Mbps ) |
|---|---|
| 1 | 44 |
| 2 | 86 |
| 4 | 165 |
| 8 | 189 |
| 16 | 340 |
| 32 | 433 |
| 64 | 615 |
| 128 | 736 |
| 256 | 813 |

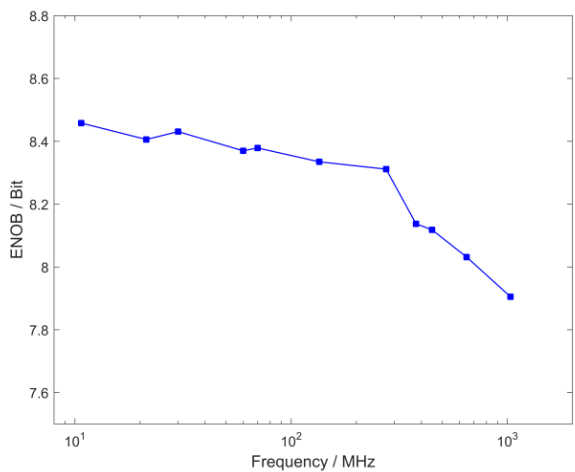

Fig. 6. Results of ENOB test.

*B. ENOB Test*

Because ENOB takes into account the effects of noise, harmonics and other factors on the sampling accuracy, it can comprehensively indicate the performance of the waveform digitizer prototype well. To evaluate the ENOB of the prototype, a series of sine wave signals with frequency ranging from 10.7MHz to 1034MHz are used. All signals pass through a bandpass filter before being connected to the prototype to reduce the impacts of the noise and harmonics of the signal generator itself. For each test, 98240 sampling points are used to calculate ENOB through Fast Fourier Transform (FFT). Test results are shown in Fig. 6, ENOB reaches 8.4 b at 10.7MHz. When the input frequency increases, the ENOB drops to 7.9 b at 1034MHz. which is much better than most commercial oscilloscopes.

## IV. CONCLUSION

In this paper, an ultra-high-speed waveform digitizer prototype based on gigabit Ethernet has been built for plasma diagnostics. The sampling rate of the prototype is as high as 5Gsps. ENOB is more than 7.9 b within 1G bandwidth. Besides, it has a large DDR3 memory cache of 4GB, all the data can be upload quickly via gigabit Ethernet interface. It can perform comprehensive physical information measurements such as time measurement, particle identification, energy measurements and so on for plasma diagnostics. Compared with commercial oscilloscopes, it has better performance and smaller volume.